\documentclass[10pt]{article}
\usepackage{ametsoc2col}
\newcommand{\myabstract}{
Characterising the stratosphere as a turbulent system, temporal fluctuations often show different correlations for different time scales as well as intermittent behaviour that cannot be captured by a single scaling exponent. In this study, the different scaling laws in the long term stratospheric variability are studied using Multifractal de-trended Fluctuation Analysis. The analysis is performed comparing four re-analysis products and different realisations of an idealised numerical model, isolating the role of topographic forcing and seasonal variability, as well as the absence of climate teleconnections and small-scale forcing. 
The Northern Hemisphere (NH) shows a transition of scaling exponents for time scales shorter than about one year, for which the variability is multifractal and scales in time with a power law corresponding to a red spectrum, to longer time scales, for which the variability is monofractal and scales in time with a power law corresponding to white noise. Southern Hemisphere (SH) variability also shows a transition at annual scales. The SH also shows a narrower dynamical range in multifractality than the NH, as seen in the generalised Hurst exponent and in the singularity spectra. The numerical integrations show that the models are able to reproduce the low-frequency variability but are not able to fully capture the shorter term variability of the stratosphere. 
}
\begin{document}

%
%
\title{\textbf{\large{Nonlinear Stratospheric Variability: Multifractal De-trended Fluctuation Analysis and Singularity Spectra}}}
%
%
\author{\textsc{Gualtiero Badin,}
				\thanks{\textit{Corresponding author address:} 
Gualtiero Badin, Institute of Oceanography,
University of Hamburg,
Bundesstrasse 53,
D-20146 Hamburg,
Germany. 
				\newline{E-mail: gualtiero.badin@uni-hamburg.de}}\\
\textit{\footnotesize{Institute of Oceanography, Universit\"at Hamburg, Germany}}
\and 
\centerline{\textsc{Daniela I.V. Domeisen}}\\
\centerline{\textit{\footnotesize{GEOMAR Helmholtz Centre for Ocean Research Kiel / University of Kiel, Germany}}}
}
%
\ifthenelse{\boolean{dc}}
{
\twocolumn[
\begin{@twocolumnfalse}
\amstitle

\begin{center}
\begin{minipage}{13.0cm}
\begin{abstract}
	\myabstract
	\newline
	\begin{center}
		\rule{38mm}{0.2mm}
	\end{center}
\end{abstract}
\end{minipage}
\end{center}
\end{@twocolumnfalse}
]
}
{
\amstitle
\begin{abstract}
\myabstract
\end{abstract}
\newpage
}

\section{Introduction}  
Recently, \citet{badindomeisen14,badindomeisen14b}, henceforth BD14a,b, explored the temporal variability of the stratosphere both in the Northern Hemisphere (NH) and in the Southern Hemisphere (SH), focussing on the chaotic nature of the variability.
The Northern and Southern Hemispheres exhibit considerably different variability. The SH winter shows less variability, being influenced by weaker topographic forcing and differential surface heating, that are the origin for planetary-scale waves, which can cause Sudden Stratospheric Warming (SSW) events in the NH. Summer variability is more similar for both hemispheres, since planetary-scale waves are inhibited from propagating into the stratosphere during this season \citep{Charney+Drazin61,Plumb89}. This yields a much stronger change in variability throughout the seasonal cycle in the NH as compared to the SH, which is still very clearly visible after removing the mean seasonal cycle. 

The analysis of the probability distribution functions (PDFs) for different variables obtained from re-analysis data as well as data generated using a dynamical core model showed that in both hemispheres the variability is characterised by a non-Gaussian distribution of the fluctuations, i.e.~the anomalies of the zonal mean zonal wind, temperature, and geopotential height. The deviation from Gaussianity in the distribution was shown to result to a major part from the higher variability in the winter season as compared to the summer season. As a consequence of this, the separation of the time series into winter and summer months revealed  distributions considerably more consistent with Gaussian distributions. Further separating the time series into specific months and for regions further poleward can yield additional deviations from Gaussianity, as e.g.~shown in \citet{YodenTaguchiNaito2002,NishizawaYoden2005}. A further speculation for the deviation from Gaussianity is the asymmetry with respect to weak and strong vortex events, as shown for the North Atlantic Oscillation in \citet{BD01}, which was however found to be a smaller factor than the seasonal cycle. 

The different variables analysed in BD14a,b showed that the NH variability is characterised by red spectra, with slopes equal to $-2$ at time scales shorter than one year and white spectra, corresponding to white noise, at longer time scales. In the SH, instead, the frequency spectra showed flatter slopes of $-1$ at time scales shorter than about one year and white spectra at longer time scales. The separation of the time series into winter and summer months showed that the spectra exhibit a $-1$ slope in the austral winter months, transitioning to slopes of $-2$ in the austral summer month. The smaller $-1$ slope in the austral winter months was explained as a signature of the longer persistence of the stratospheric polar vortex in the SH as compared to the NH.

These characteristics of stratospheric variability also give insight into possible intrinsic predictability of the stratosphere. Within the stratosphere, it is generally assumed that  predictability on seasonal or longer timescales is low, i.e.~that "a major source of stratospheric variability (and thus predictability) is located in the troposphere below and not in the stratosphere itself" \citep{PolvaniWaugh2004}. It has however been shown that teleconnections, e.g.~from the tropical stratosphere in terms of the Quasi-Biennial Oscillation (QBO), e.g.~\citet{Garfinkeletal2012,Scaifeetal2014}, and from the tropical troposphere in terms of El Ni\~no Southern Oscillation (ENSO), e.g.~\citet{InesonScaife2008,ButlerPolvani2011,Domeisenetal2015} and the Madden - Julian Oscillation (MJO), e.g.~\citet{Garfinkeletal2014}, can have an influence on the extratropical stratosphere, indicating that some or all of these teleconnections may add seasonal predictability. In addition, the extratropical troposphere and the extratropical land/ocean surface may also have a longer term influence on the stratosphere, e.g.~regarding snow cover \citep{CohenEntekhabi1999} or sea ice \citep{SunDeserTomas2015}. Recently, \citet{Stockdaleetal2015} have shown through model initial conditions that even the extratropical atmosphere may have intrinsic predictability on seasonal timescales. Predictability is thus often defined from comprehensive models that represent the predictive processes to a certain degree in comparison to reanalysis and observational data that is used to initialise the model. 

This indicates that a realistic representation of these predictive processes, which is often obtained through nudging of the model to observations, can  improve predictability. This study, in contrast, aims at identifying the dominant variability and possible predictability within the stratosphere from the analysis of time series of the extratropical stratosphere. The comparison between reanalysis and simplified models in this study will ideally allow for conclusions about the inherent predictability of the extratropical stratosphere that may arise through mechanisms other than teleconnections from processes that may not be represented in the simplified model.

In particular, the behaviour of both the PDFs and the frequency spectra of the fluctuations suggests the potential existence of different scaling regimes of the fluctuations, i.e.~a multifractal variability of the system, at different time scales. Loosely speaking, the identification of multifractal variability in the stratosphere could imply different scaling regimes at different time scales, e.g. through the presence of different scaling invariance relations at different temporal scales. Plus, multifractality is related to the intermittency of the system. Multifractal behaviour is common in geophysical fluids, and the search for the existence of scaling regimes in atmospheric dynamics has a long history (e.g.~\citet{LovejoySchertzer13} and references therein).  In this study, we search for a possible multifractal structure of the stratospheric variability for different variables and re-analyses, in different locations of the stratosphere, as well as for different realisations of a numerical model, based on the GFDL three-dimensional atmospheric spectral dynamical core model in an idealised set-up. Each realisation of the model is aimed at separating the importance of the seasonal forcing and of the topographic forcing in the stratospheric variability. 

The technique to characterise the multifractal variability of the system from time series will be based on the Multifractal de-trended Fluctuation Analysis (MF-DFA) \citep{kantelhardt_etal01,kantelhardt_etal02,kantelhardt2012fractal}. Loosely speaking, the MF-DFA analyses the behaviour of fluctuation functions at different time scales, as a function of a continuous parameter that defines different moments. The power laws so derived allow for a characterisation of an entire spectrum of fractal dimensions of the system. MF-DFA and its modifications have a long history of applications, and they were employed for the climate system to study for example the long-term variability of climate \citep{ashkenazy03}, cloud patterns \citep{arrault97}, precipitation and river runoff \citep{kantelhardt03}, wind speeds \citep{govindan04} and Arctic sea ice persistence \citep{agarwal12}. The identification of a multifractal nature for the scaling laws of the atmosphere allows for a characterisation of the dynamics and of the intermittency of the fluctuations, for example through the use of random multiplicative cascades \citep{schertzerlovejoy87,schertzerlovejoy97,schertzeretal97,LovejoySchertzer13} that do not require the assumption of the existence of low dimensional chaotic variability in the system, as discussed in BD14a,b for the case of the stratosphere.

To the authors' knowledge, this is the first time that this technique has been applied to the stratospheric circulation. It will give insights into the different degrees of nonlinearity of the system (as explained in the next Section) in the NH and SH, as well as a possible way to assess to what extent idealised models reproduce the variability of the stratosphere. 


\section{Theory}

\subsection{Scaling laws for turbulent fluctuations}

Consider a time series $x_i$, and the fluctuations $\Delta x_i =  x_i - < x > $, where $< x > = (1/N) \sum_{i=1}^N x_i$ is the mean. The autocorrelation function for the $\Delta x_i$ scales as 
\begin{equation}
C(s) = <  \Delta x_i   \Delta x_{i+s}  >   ~,
\label{eq:s0}
\end{equation} 
where $s$ is a scale of the fluctuations, which in the case of a time series is a time interval $\Delta t$. Considering the mean correlation time $\tau_s = \int_0^N C(s) ds$, if $0 < \gamma <1$,  $\tau_s$ diverges as $N \rightarrow \infty$, and the $\Delta x_i$ are said to be long-range correlated and 
In the case of long term persistence, the autocorrelation function for the $\Delta x_i$ scales as 
\begin{equation}
C(s)  \sim s^{- \gamma} ~,
\label{eq:s0_1}
\end{equation} 
holds; if $ \gamma \ge 1$,  $\tau_s$ converges as $N \rightarrow \infty$, and the $\Delta x_i$ are said to be short-range correlated; if instead $C(s) = 0$ for a certain $s>0$, the $\Delta x_i$ are said to be uncorrelated. 
Very often, natural systems, and especially turbulent systems, are not characterised by a single exponent $\gamma$, with time series showing different correlations for different scales $s$. Furthermore, the time fluctuations can exhibit intermittency, which cannot be represented by a single scaling exponent. In order to take these effects into account, it is possible to introduce the structure functions
\begin{equation}
< \left( \Delta x_i   \Delta x_{i+s} \right)^q >  \sim s^{\zeta (q)} ~.
\label{eq:s1}
\end{equation} 

The function $\zeta (q)$ is called structure function exponent and can be expressed as the sum of a linear and a nonlinear part
\begin{equation}
\zeta (q) = q H - K(q) ~,
\label{eq:s3}
\end{equation}
where $H$ is the Hurst exponent \citep{Hurst51} and the nonlinear part $K(q)$ is called the {\it moment scaling exponent} \citep{LovejoySchertzer13}. In the assumption of quasi-Gaussianity, the classical Hurst exponent has the properties that if $0 < H < 0.5$, the time series $x_i$ is considered to be long-range anti-correlated. If $H = 0.5$ or $H > 0.5$, the time series is considered to be uncorrelated or long-range correlated, respectively. $K(q)$ is a convex function related to the $q$-moment of the turbulent fluxes responsible for the intermittency and multifractality of the system. 
Assuming that the intermittency of the system is created by a multiplicative cascade, the exponent $K(q)$ takes the universal form \citep{schertzerlovejoy87}
\begin{equation}
K(q) = \frac{C_1}{\left( \alpha_1 - 1 \right)} \left( q^{\alpha_1} - q \right)~,
\label{eq:s5}
\end{equation}
for $0 \le \alpha_1 \le 2$, and $K(q)=C_1 q \log q$ for $\alpha_1 =1$
Under this assumption, the study of the multifractality of the system is reduced to the determination of the constants $C_1$, i.e. che codimension of the system, and $\alpha_1$, which is also called the Levy exponent. 
Limiting cases of (\ref{eq:s3}) are the trivial one in which the system is non-intermittent, i.e. $K(q)=0$;
the case in which $K(q)$ is linear in $q$, that corresponds to the case in which the time series is said to be monofractal. A particular model that applies in this case is given by the beta model \citep{frischetal78}, where $\alpha_1=0$ and
\begin{equation}
K(q) = C_1 (q-1)~;
\label{eq:s3b}
\end{equation}
finally, the case $\alpha_1=2$ corresponds to a lognormal multifractal for which
\begin{equation}
K(q) = C_1 q(q-1)~.
\label{eq:s3c}
\end{equation}
The search for scaling laws characterising the fluctuations of the system becomes thus related to the characterisation of the structure function exponent. 
One way to characterise the  structure function exponent is through the Multifractal de-trended Fluctuation Analysis (MF-DFA) \citep{kantelhardt_etal01,kantelhardt_etal02,kantelhardt2012fractal}. In the following we will expose how to perform the MF-DFA analysis and how to relate it to equations (\ref{eq:s1})-(\ref{eq:s3}).

\subsection{Multifractal de-trended Fluctuation Analysis}

The MF-DFA proceeds in several steps, here reported following \citet{kantelhardt_etal02}: 
\begin{itemize}
\item Define the cumulative anomalies $y_i$, defined as
\begin{equation}
y(i) := \sum_{j=1}^i \Delta x_j~,
\label{eq:6}
\end{equation} 
where $i=1,..., N$.

\item Divide the profile $y(i)$ into $N_s= N/ s$ non-overlapping segments of size $s$. 

\item For each segment, calculate the local trend $y_r (i)$ using a least-square polynomial fit. Because the $N_s$ segments will not necessarily cover the entire time series, this process will be performed twice, once forward in time, starting from the beginning of the time series, and once backward, starting from the end of the time series. The process will thus be performed over $2 N_{s}$ segments. Once the local trend for each of the $2 N_s$ segments is calculated, determine the variance with respect to the local trend
\begin{equation}
Var (r, s) := \frac{1}{s} \sum_{i=1}^{s} \left[ y \left( ( r -1 ) s +i \right) - y_r (i) \right]^2~,
\label{eq:7}
\end{equation}  
where $r=1,...,2N_{s}$.
\item From (\ref{eq:7}) it is possible to define the generalised fluctuation function as
\begin{equation}
F_q (s) = \left[ \frac{1}{2 N_{s}} \sum_{r=1}^{2 N_{s}} \left[ Var(r,s) \right]^{\frac{q}{2}} \right]^\frac{1}{q}~,
\label{eq:8}
\end{equation}  
where $q$ is a continuous parameter that defines different moments. For $q=2$, the usual de-trended fluctuation analysis (DFA) is retrieved \citep{peng_et_al94}. Because (\ref{eq:8}) is singular for $q \rightarrow 0$, it is possible to define
\begin{equation}
F_q (s) = \exp \left[ \frac{1}{4 N_{s}} \sum_{r=1}^{2 N_{s}} \log \left[ Var(r,s) \right] \right]~,
\label{eq:8b}
\end{equation}  
for $q=0$.

\end{itemize}

The aim of the MF-DFA is to study the generalised fluctuation function as the choices of the continuous parameter $q$, the length of the time segment $s$ and the degree of the polynomial fit are varied. In particular, if the series $x_i$ are long-range correlated, (\ref{eq:8}) shows a power law dependence
\begin{equation}
F_q (s) \sim s^{h(q)}~.
\label{eq:9}
\end{equation} 

In (\ref{eq:9}), the scaling exponent $h(q)$ is called the {\it generalised Hurst exponent}. It should be noted that $h(q)$ is related to the structure function exponent as 
\begin{equation}
h(q) = 1 + \frac{ \zeta (q)}{q}~ .
 \label{eq:s4}
\end{equation} 

\subsection{Generalised Dimensions and Singularity Spectra} 

To relate the results from the MF-DFA analysis and the generalised dimensions $D_q$, following \citet{kantelhardt_etal02} consider the series $x_i$ as stationary and normalised, i.e. $\sum^{N}_{j=1} x_j = 1$. Then the calculation of the variance (\ref{eq:7}) does not need de-trending and the variance simplifies to
\begin{equation}
Var (r,s) = \left[ y \left( ( r -1 ) s \right) - y (rs) \right]^2~.
\label{eq:10}
\end{equation} 
With (\ref{eq:10}), the fluctuation function (\ref{eq:8}) reduces to
\begin{equation}
F_q (s) = \left[ \frac{1}{2 N_{s}} \sum_{r=1}^{2 N_{s}} \left| y \left( ( r -1 ) s \right) - y (rs) \right|^{\frac{q}{2}} \right]^\frac{1}{q} ~.
\label{eq:11}
\end{equation} 
Assuming that the length of the time series is a multiple of $s$, and with the help of (\ref{eq:9}), (\ref{eq:11}) can be expressed as
\begin{equation}
F_q (s)  \sim s^{q h(q) -1} = s^{\tau(q)}~,
\label{eq:12}
\end{equation} 
where
\begin{equation}
\tau(q) = q h(q) -1~,
\label{eq:13}
\end{equation} 
and thus, in terms of the structure function exponent,
\begin{equation}
\tau(q) = q +\zeta(q)-1=q \left( H+1 \right) - \left( K (q) +1 \right) ~.
\label{eq:13a}
\end{equation} 
Note that $\tau(q)$ should not be confused with the mean correlation time $\tau_s$ calculated from Equation \eqref{eq:s0}.
The term 
\begin{equation}
\mu(r,s) = \sum_{k=(r-1)s+1}^{rs} x_k = y \left( [ r -1 ] s \right) - y (rs)~, 
\label{eq:14}
\end{equation} 
is the measure of the system.
From the measure it is possible to derive the quantity 
\begin{equation}
I (s) = \sum_{r=1}^{N/s} \left| \mu(r,s) \right|^q \sim s^{\tau (q)}~.
\label{eq:15}
\end{equation} 
that should be compared with Equation (4) of BD14a.
%
The different dimensions are defined as   
\begin{equation}
D_{q} = \frac{\tau(q)}{q-1}  ~,
\label{eq:2}
\end{equation}
and thus, in terms of the structure function exponent,
\begin{equation}
D_{q} = 1 + \frac{\zeta (q)}{q-1} = 1 + H \frac{q- \frac{ K (q) }{ H }}{q-1}~.
\label{eq:2bis}
\end{equation}
A complete derivation of (\ref{eq:2}) can be found in \citet{parisifrisch85} and \citet{halseyetal86}. The case $q=0$ yields the box couting dimension $D_0$, $q=1$ yields the information (or Shannon) dimension $D_{1}$, $q=2$ yields the correlation dimension $D_2$ \citep{Grassberger+Procaccia1983a, Grassberger+Procaccia1983b}. It should be noted that, despite the meaning of the special cases $q=0,1,2$, $q$ is to be intended as a continuous index. In general, it can be shown that $D_{q_1} \le D_{q_2}$ if $q_1 > q_2$. If the equality sign holds, the system is called monofractal, otherwise it is called multifractal. 

In the multifractal case, the system is characterised by a spectrum of values of dimensions. To determine this spectrum, it is possible to introduce a Legendre transform of $\tau (q)$ defined as
\begin{eqnarray}
\alpha = \frac{d \tau}{d q}~, ~f( \alpha ) = q \alpha - \tau (q)~, ~ q = \frac{d f ( \alpha )}{d \alpha}~.
\label{eq:16}
\end{eqnarray} 
With (\ref{eq:16}), (\ref{eq:2}) can be written as
\begin{equation}
D_{q} = \frac{1}{q-1} \left[ q \alpha (q) - f \left( \alpha (q) \right) \right]~.
\label{eq:17}
\end{equation}
Equation (\ref{eq:17}) shows that $D_q$ and $f(\alpha)$ give the same amount of information, with the quantity $f ( \alpha )$ expressing the distribution of dimensions of the set upon which the singularities of strength $\alpha$ may lie. The study of multifractality via the distribution of the singularity exponent $\alpha$ offers however an elegant geometric interpretation in which $\alpha$ is the H\"older exponent of the cascade \citep{parisifrisch85}. Loosely speaking, the larger the singularity, the larger the growth of fluctuations at a certain time ratio of the cascade process. In infinite dimensional systems, $f(\alpha)$ is linked to the codimension and can thus be used for a stochastic representation of the system (see e.g.~\citet{LovejoySchertzer13}). 


\subsection{Guidelines for the Application of the Theory}

The theory summarised above yields as results a number of functions: the generalised Hurst exponent $h(q)$, the fluctuation functions $F_q$ and the singularity spectra $f ( \alpha )$. How does one interpret the results of a time series analysis based on these parameters?

First of all, note that (\ref{eq:9}) might fail for a time series both for small values of the time interval $s$ as well as for very large values of $s$, where the number of segments $N_{s}$ becomes small. 

If the system is non-intermittent, i.e. for $K=0$, equation (\ref{eq:s4}) yields $h=1+H$, so that $h$ is a constant. 
For the monofractal beta model (\ref{eq:s3b}), one has instead $h(q) = H - C_1 \left( 1- 1/q \right)$, which diverges for $q \rightarrow 0$. Finally, for the log-normal cascade (\ref{eq:s3c}), $h(q)=(1-H-C_1)+C_1 q$.

The generalised Hurst exponent $h(q)$ is also connected to the power spectrum of the time series through the relation $h(2)=(1+\beta)/2$, where $\beta$ is the decay rate of the power spectrum of frequency $f$, so that $S(f) \propto f^{- \beta}$. The quantity $h(2)$ thus gives  an indication of the kind of variability of the system: for white noise, $\beta = 0$ and thus $h(2) = 1 / 2$, while for red noise $\beta = 2$ and $h(2) = 3/2$.
 
The value of $h(2)$ is connected to the slope of the fluctuation function through (\ref{eq:9}) for $q=2$. Through the dependence of $F_2$ on $s$ it is possible to study if for different interval lengths the system variability tends to variability with a different color, i.e. with a different frequency spectrum. 

If the exponent $\tau (q)$ does not depend on $q$ the system will be monofractal, while a dependence on $q$ will indicate that the system is multifractal.
One should note that for positive $q$, the segments with large variance, corresponding to large deviations from the polynomial fit, will dominate  $F_q$. In this case, $h(q)$ describes  the scaling behaviour of the segments with large fluctuations. For negative $q$, $h(q)$ instead describes the scaling behaviour of the segments with small fluctuations.

Finally, the width of the singularity spectrum $f(\alpha)$ is linked to the degree of nonlinearity of the system. Consider for example once again the non-intermittent system $K=0$. In this case, from (\ref{eq:16}) $\alpha=H+1$ and $f(\alpha)=0$, so that the singularity spectrum has zero amplitude. For the beta model (\ref{eq:s3b}), one has $\alpha=H+1-C_1,~f(\alpha)=1$, i.e. the singularity spectrum is reduced to a single value, in agreement with the monofractal nature of the model. For the log-normal cascade (\ref{eq:s3c}), $\alpha=(1-H-C_1)+2 C_1 q$ and $f(\alpha)=C_1 q^2 + 1$.
The relationship between $f(\alpha)$ and $F_q$, determined by the scaling coefficient $h(q)$, shows that a system dominated by large differences in $F_q$ determined by large and small fluctuations will be characterised by a larger singularity spectrum. 

In the following, the theory will be applied to  stratospheric variability both in the NH and in the SH. Calculations are done varying the value of the moment $q$ between $-20$ and $20$, which was used for example in the studies by \citet{kantelhardt_etal02}, \citet{kantelhardt03} and \citet{agarwal12}, using $200$ intervals.  Because the choice of the interval $[-20~20]$ is arbitrary, it has been checked that the different quantities do not vary for large or small values of $q$. Most of the variability is indeed in the interval $-3 \le q \le 3$, possibly due to the phenomenon of multifractal phase transitions \citep{schertzerlovejoy92}. Outside of this interval, the different diagnosed quantities might be affected by spurious multifractality. To check for spurious multifractality we checked the convergence of the measure $\Delta h = h(- \infty)-h(+\infty)$ \citep{schumann_kantelhardt11}. Following \citet{schumann_kantelhardt11}, the measure $\Delta h_{20} = h(-20)-h(20)$ has been considered here as a criterion.  

The presence of spurious multifractality due to the finite length of the time series has been investigated through an analysis of the results after a random shuffling of the original time series. Random reshuffling of the time series should yield white noise, with the results differing from a Gaussian distribution due to the effect of the finite length of the time series. 
To test for the effect of autocorrelations of the time series, we have also performed a reshuffling of the time series after their division into blocks with length determined through mutual information (\citet{Fraser+Swinney86}, see also BD14a,b), as well as the reshuffling of blocks without the reshuffling of the elements within the boxes. Both tests confirm the lack of spurious multifractality, with however not shown, in order to focus on the predictability of the system focussing on the entire time series. For a complete analysis of spurious multifractality with different reshuffling of the time series, see \citet{schumann_kantelhardt11}.

The results have been tested for polynomial fitting ranging from order one (linear fitting) to order three. In all cases, the results hold, suggesting a stationarity in the variability.



\section{Data}

The examined variables are daily and zonal mean zonal wind, temperature and geopotential height at 10$\,$hPa and at the grid point closest to 60$^\circ$ N and 60$^\circ$ S (the same as in BD14a,b). Zonal mean zonal wind at 60$^\circ$ N and 10hPa as well as  temperature averaged over an extratropical latitude band are commonly used to define polar vortex strength and stratospheric sudden warming events \citep{McInturff78}.

\subsection{Re-analysis Data}

The re-analysis data sets examined in this study are: the ERAinterim re-analysis \citep{deeetal11} for the period  1 Jan 1980 - 31 Dec 2011, yielding a total of 11,688 daily data points; the ERA40 re-analysis \citep{uppalaetal05}, for the period  1 Jan 1958 - 31 Dec 2001, yielding a total of 16,071 data points; the NCEP re-analysis \citep{kalnayetal96}, for the period  1 Jan 1948 - 31 Dec 2012, for a total of 23,742 points; and the NCEP-DOE AMIP-II re-analysis (R-2) data set \citep{kanamitsuetal02}, hereafter referred to as NCEP2,  for the period 1 Jan 1979 - 31 Dec 2012, yielding 12,419 data points. The seasonal cycle has been removed from each re-analysis time series by subtracting the climatological seasonal cycle for leap years and non-leap years separately. A qualitative description of the variability of the different data sets is reported in BD14a,b.

\subsection{Model Integrations}

The model used for this study is the GFDL 3-dimensional atmospheric spectral dynamical core model at T42 resolution on 40  hybrid $\sigma$-pressure levels up to 0.02 hPa (with 28 levels above 200hPa) and a sponge layer starting above 0.8 hPa. The model uses a Newtonian relaxation to a zonal mean equilibrium temperature profile based on \citet{heldsuarez94} in the \citet{polvanikushner02} setup (using $\gamma$ = 4 K / km and $\epsilon$ = -10 K). The relaxation time scale is 40 days in the stratosphere. Zonal wave-2 topography of height 3000 m is used to force stratospheric variability as defined in \citet{gerberpolvani09}. 

These model runs, especially the run with topography, are designed to give a variability comparable to the real atmosphere, especially in terms of the number and frequency of stratospheric sudden warming (SSW) events \citep{gerberpolvani09} as defined by the variability at 60N and 10hPa. While comprehensive climate models nowadays tend to exhibit resolutions greater than T42, the goal of this study is to investigate if this similar variability holds up to the tests available in time series analysis. 

Three model runs are performed. Model I is forced with a 360-day seasonal cycle (as defined in \citet{kushnerpolvani06}) and wave-2 topography, Model II is run without a seasonal cycle, i.e.~in perpetual winter conditions for the analysed hemisphere and including topography, while Model III is run without a seasonal cycle and without topography and thus lacks the main drivers of large-scale stratospheric variability at the considered pressure level. The specific run performed for Model I is further documented in \citet{sheshadrietal13}, while the run performed for Model II is described in \citet{gerberpolvani09} (their run 9), except here run on hybrid levels (instead of $\sigma$ levels). See BD14a for a qualitative description of the variability of the Model I and II integrations.

Model I is integrated for 13,000 days, and the last 33 yearly cycles are used for the analysis. The seasonal cycle is removed by subtracting the 360-day climatological seasonal cycle from each model year. Model II is integrated for 29,900 days. Model III is integrated for 9,800 days. The model output was saved as daily mean values for all model runs. 




\section{Results}

\subsection{Fluctuation Functions}

\subsubsection{Northern Hemisphere}

The dependence of the fluctuation functions $F_q$ on the length of the time interval $s$ is shown in Figure \ref{fig1}a for the NH zonal mean zonal wind from the ERAInterim re-analysis. The grey curves show the fluctuation functions for $q$ varying from $q=-3$ (lower line) to $q=3$ (upper line). The line corresponding to $F_2$, i.e.~$q=2$, is printed in black. Starting the analysis from $F_2$ and for small values of $s$, $F_2$ has a slope that is similarly or less steep than $3/2$, corresponding to red noise. For $s \ge 365$ days, the slope is flatter and matches the slope of $1/2$, corresponding to white noise. This result is an indication that for time scales shorter than about one year, the variability is almost Brownian, i.e.~corresponding to red noise, while for longer time scales it exhibits a transition to white noise. This is in agreement with the slopes of the frequency spectra in the NH found by BD14a that show $\beta=2$ spectral slopes for time scales shorter than 
 one year and  white spectra for longer time scales. 

This result  also characterises the predictability of the system, suggesting that the longer term behaviour of the NH stratosphere is dominated by random, white fluctuations, indicating the limitations in year-to-year predictability. However, a change in slope can be observed around a time scale of 800 days ($\sim 26$ months), while the slope between the annual time scale and the 26 month time scale is close to 1/2, i.e.~white noise. A steeper slope could imply a possible increase in predictability (or at least a deviation from white noise) for the time scale around the period of the QBO. The time series is  however rather too short to be able to infer this with certainty. In addition, like other teleconnections, the QBO influences the extratropical stratosphere (e.g.~through the Holton-Tan mechanism) in a different fashion and strength throughout the annual cycle, similar to ENSO, which influences the extratropical stratosphere predominantly in winter. Therefore, the QBO, while varying on longer time scales, will in addition project onto the annual cycle variability of the extratropical stratosphere. 

The analysis of other moments, i.e.~other values of $q$, shows that for $s \le 1$ year, the different curves are well separated and the system appears thus to be multifractal. Higher moments, e.g. $q < -3$ or $q > 3$ (not shown), collapse onto the curves representing these two values, suggesting these limits for the multifractal behaviour of the system, in agreement with the idea of multifractal phase transitions \citep{schertzerlovejoy92}.

For time scales of around one year or longer, the curves collapse onto the same curve, corresponding to monofractal behaviour at longer time scales. Keeping in mind that the seasonal cycle has been removed from the time series, the threshold of one year for the passage from a multifractal to a monofractal variability is likely associated with the transition from a weaker variability in the summer months to a stronger variability during the winter season (BD14a,b). For time scales between around half a year to a year, the curves for larger values of q tend towards a slope of $1/2$, while the curves for negative values of q retain a steeper slope for longer as the time scales increase.  This suggests, since negative values of $q$ correspond to segments with smaller fluctuations, i.e.~summer variability, that these segments tend to be more predictable on time scales of several weeks to months than segments with large fluctuations, i.e.~winter variability, which is not surprising given the smaller deviations from climatology of the summer months as compared to the winter months. 
In comparison, in the troposphere a transition is found at the much shorter time scale of around 10 days, corresponding to the transition between weather and macroweather \citep{lovejoyetal2010,lovejoyschertzer11}.

In order to test for the presence of spurious multifractality due to the finite length of the time series, an analysis of the fluctuation functions for the NH zonal mean zonal wind from the ERAInterim re-analysis after a random shuffling of the original time series has been performed (Figure \ref{fig1}c). The results indicate that with the random shuffling of the time series, the curves collapse onto the same curve, with a slope of $1/2$, corresponding to white noise. Only a small spread between the lines is present due to the finite length of the time series, showing a clear distinction from the spread in Figure \ref{fig1}a.


\subsubsection{Southern Hemisphere}

The analysis of the fluctuation function for the SH zonal mean zonal wind from the ERAInterim re-analysis (Figure \ref{fig1}b) again shows a transition from a slope of less than $3/2$ to $1/2$ as time scales increase. This is in agreement with the slope of the frequency spectrum in the SH found by BD14b, which shows slopes of $\beta=1$ for shorter time scales, corresponding to $h(2)=1$. For longer time scales, the slope of $F_2$ again approaches  $1/2$, corresponding to white noise. The transition to a smaller slope is considerably smoother as compared to the NH, reflecting the smaller difference between summer and winter variability in the SH. The analysis of higher moments again suggests a multifractal behaviour of the system, but with a  smaller spread between the lines representing the different moments as compared to the NH. 

\subsubsection{Model Integrations}

\begin{figure}[t]
  \noindent\includegraphics[width=30pc,angle=0]{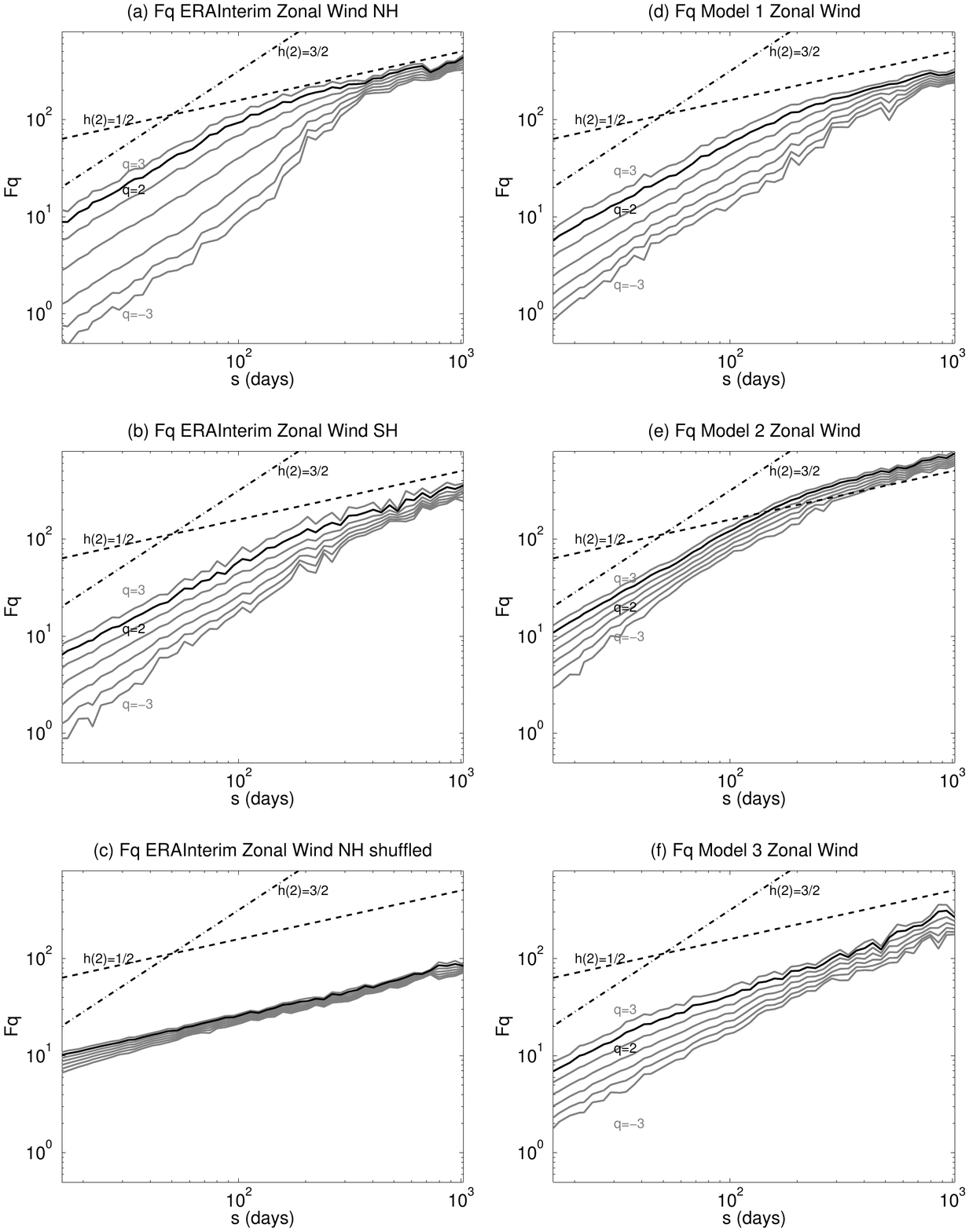}\\
  \caption{Fluctuation functions $F_q (s)$ for (a) the Northern Hemisphere and (b) the Southern Hemisphere zonal mean zonal wind from ERAInterim re-analysis. Panel (c) represents the same analysis as (a), but after a random shuffling of the data. (d) Model I, (e) Model II and (f) Model III zonal mean zonal wind. The gray curves show the fluctuation functions for $q$ varying from $q=-3$ (lower gray curve) to $q=3$ (upper gray curve). The curve corresponding to $q=2$ is printed in black. The dashed line indicates the slope corresponding to $h(2)=1/2$. The dot-dashed line corresponds to the slope $h(2)=3/2$. }\label{fig1}
\end{figure}

How well is the statistical behaviour of the stratosphere, as represented by the fluctuation functions, reproduced in the different model integrations? Figure \ref{fig1}d shows the fluctuation function for the zonal mean zonal wind for the Model I integration, i.e.~for the model forced  by both a seasonal cycle and topography. The second moment, $F_2$,  qualitatively reproduces the slopes observed in the NH for the ERAInterim re-analysis (Figure \ref{fig1}a), with a transition between a slope of $3/2$ for shorter time scales to $1/2$ for longer time scales. The analysis of higher moments, i.e.~different values of $q$, however shows a departure of the behaviour between the Model I integration and the ERAInterim re-analysis: The spread between the lines corresponding to different moments is less pronounced in the numerical model and  instead resembles the spread between the different modes of the SH (Figure \ref{fig1}b). This result suggests that the model might underestimate the variability of the system at time scales shorter than about a year. This may seem indicative of the longer decorrelation times in idealised models as compared to the re-analysis, as e.g.~documented in \citet{Chan:2009fv} and \citet{Gerber:2008ug}. This problem has however also been observed for more comprehensive models \citep{GerberPolvaniAncukiewicz2008} with higher resolution, and long decorrelation times are  less of a problem in the stratosphere as compared to the troposphere. In the model runs used for this study, stratospheric decorrelation times are close to 40 days, which corresponds to the radiative relaxation time scale. This finding may therefore rather be indicative of missing sub-grid scale processes due to the comparably coarse resolution, such as for example gravity waves.  In terms of dynamical systems analysis, the result indicates that the model might have a more uniform measure along the attractor than the re-analysis. It should also be noted that the convergence of the curves for the different moments occurs at longer time scales as compared to the ERAInterim re-analysis.

The Model II integration (with topography but without a seasonal cycle) shows a similar behaviour, with an even narrower spread between the curves for the different moments than observed in the SH ERAInterim re-analysis (Figure \ref{fig1}e), implying a different variability in the model than in the re-analysis, at least in statistical terms. Comparing the results for Model II and Model I indicates that the inclusion of a seasonal cycle is responsible for part of the spread between the curves for the different moments. Note that Model II also shows a change in the slope of $F_q$ at time scales around $\sim 1$ year, due to the fact that, despite the absence of a seasonal cycle, this integration still exhibits significant long term variability: Stratospheric sudden warming events occur in this model run at a frequency of around 300 - 500 days (depending on the criterion used to identify the events). This long term variability is reflected in a change of slope of $F_q$ associated with the cumulative sum of the fluctuations.

Finally, Model III shows slopes $> 1/2$ and $\ll 3/2$, which indicates that the internal variability of the model is close to (but not completely matching) white noise. This would be expected, as the model run does not include the external forcing by topography which forces the stronger variability observed in the other model runs. As the model produces waves and therefore stratospheric variability internally, a match with pure white noise would not be expected either. 

Comparing all model runs indicates that the collapse onto a single curve for longer time scales, as seen in ERAInterim and Model I is predominantly due to the inclusion of a seasonal cycle and the resulting change in variability on semi-annual to annual timescales. 


\subsection{Generalised Hurst Exponent and Scaling Functions}   

\begin{figure}[t]
  \noindent\includegraphics[width=30pc,angle=0]{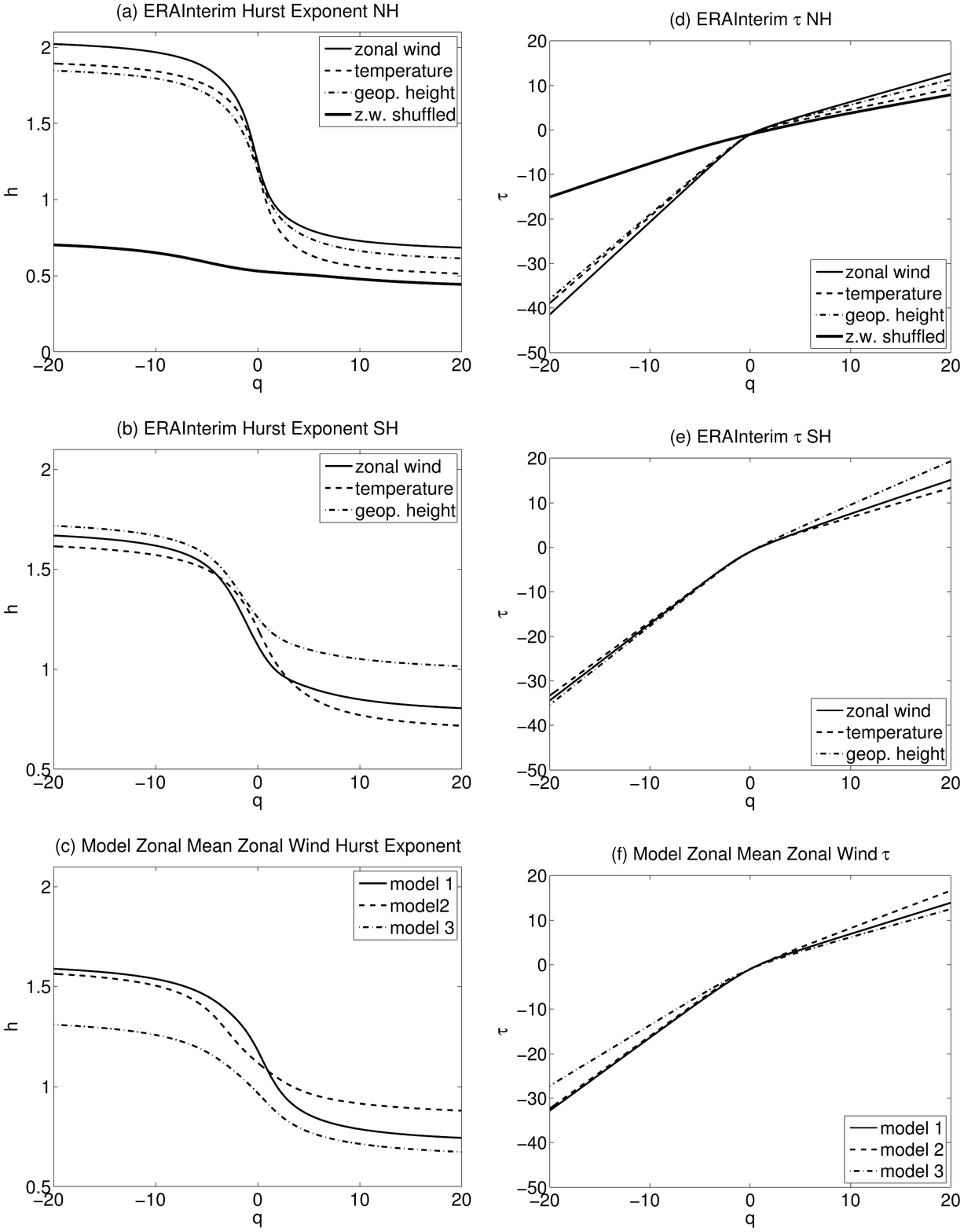}\\
  \caption{Left panels: generalised Hurst exponents $h(q)$ from ERAInterim re-analysis for  (a) the Northern Hemisphere and (b) the Southern Hemisphere for zonal mean zonal wind (solid lines), zonal mean temperature (dashed lines) and zonal mean geopotential height (dot-dashed lines). (c) generalised Hurst exponents $h(q)$ for the zonal mean zonal wind from Model I (solid line), Model II (dashed line) and Model III (dot-dashed line). Right panels: Function $\tau(q)$ from ERAInterim re-analysis for  (d) the Northern and (e) the Southern Hemispheres and for zonal mean zonal wind (solid lines), zonal mean temperature (dashed lines) and zonal mean geopotential height (dot-dashed lines). (f) Function $\tau(q)$ for the zonal mean zonal wind from Model I (solid line), Model II (dashed line) and Model III (dot-dashed line). The bold black lines in (a) and (d) show the analysis of the Northern Hemisphere zonal mean zonal wind after a random shuffling of the time series.}\label{fig3}
\end{figure}

The generalised Hurst exponents are calculated as the linear approximation of the entire curve $F_q$. Figure \ref{fig3}a shows the value of $h(q)$ for $-20 \le q \le 20$ for the ERAInterim re-analysis. The results show large differences between the values of $h(q)$ as a function of $q$. All the variables and data sets indicate that $H-1$ shows a transition from being positive at short time scales ($q<0$) to being negative at longer time scales ($q>0$). The three different variables show values of the dynamic range $\Delta h_{20} = h(-20)-h(20)$, with $\Delta h_{20}$(zonal wind)$= 1.3$, $\Delta h_{20}$(temperature)$= 1.4$, $\Delta h_{20}$(geopotential height)$ = 1.2$, giving thus a mean value $\Delta h_{20} = 1.3 \pm 0.1$. The similar values assumed by $\Delta h_{20}$ for the different variables confirm a multifractal variability of the system with coherent statistics between the different variables. All the variables and data sets show $\Delta h_{20} \sim O(1)$, ruling out the presence of spurious multifractality \citep{schumann_kantelhardt11}.

Furthermore, for all variables the dynamic range $h(-20)-h(0)$ is larger than $h(0)-h(20)$. Because the negative values of $q$ are associated with small fluctuations, it is these that dominate the variability of the system when the entire length of the time series is considered. 

The analysis of $h(q)$ for the NH zonal mean zonal wind from the ERAInterim re-analysis after a random shuffling of the original time series (Figure \ref{fig3}a, thick black line) shows that $H-1$ assumes negative values for all $q$. The distribution of $h(q)$ is close to 0.5 for all $q$, corresponding to white noise, with a deviation from 0.5 at $|q|>>0$ related to the finite length of the time series. For the shuffled time series, $\Delta h_{20}$(shuffled zonal wind)$= 0.26$, with $h(-20)-h(0)$ larger than $h(0)-h(20)$.

The analysis of the generalised Hurst exponents for the same variables and data sets, but for the SH, shows a narrower dynamical range (Figure \ref{fig3}b). In more detail, for the SH, $\Delta h_{20}$(zonal wind)$ = 0.9$, $\Delta h_{20}$(temperature)$ = 0.9$, $\Delta h_{20}$(geopotential height)$ = 0.7$, giving thus a mean value $\Delta h_{20} = 0.8 \pm 0.1$. Note that, even if the variables are connected through thermal wind balance,  geopotential height is expected to have a different variability than the zonal wind and  temperature, as noted by BD14b, as the former represents the entire integrated column below 10hPa, while wind and especially temperature are likely to be more locally controlled by the location of the polar vortex and local mixing.

Finally, the analysis for the zonal mean zonal wind for the different model runs shows that the generalised Hurst exponents have dynamical ranges $\Delta h_{20}$(Model I)$= 0.9$,~$\Delta h_{20} $(Model II)$ = 0.7$,~$\Delta h_{20}$(Model III)$ = 0.6$ (Figure \ref{fig3}c). The values of the dynamical ranges for all model runs are comparable or smaller than the values for the SH in ERAInterim. This would be expected for Model III, which exhibits no external forcing, while Model I should exhibit a dynamical range more similar to ERAInterim in the NH due to the presence of both topography and the seasonal cycle. However, the dynamical range for Model I remains similar to the SH in ERAInterim. 
However, when comparing shorter term fluctuations (negative values of $q$) and longer term fluctuations (positive values of $q$) between the models and the re-analysis, Model I tends to represent longer term (i.e.~related to large-scale) fluctuations well, with values similar to the NH in ERAInterim. Model II, with no seasonal cycle, exhibits even stronger large-scale fluctuations, as winter variability is not limited by the transition to summer. The relatively good representation of large-scale fluctuations in the models, at least at the qualitative level, confirms the purpose of the modelling exercise, with the goal of representing the long term stratospheric variability through the inclusion of large-scale forcing in the form of topography. The shorter term (i.e.~related to small-scale) variability is however considerably underestimated in the models as compared to the re-analysis. 

The multifractality of the system is determined by the behaviour of the function $\tau(q)$ defined in (\ref{eq:13}). In a monofractal system, $\tau(q)$ would thus be a straight line, with $h(q)$ being a constant slope. In a multifractal system, $\tau(q)$ instead depends on $q$. Figure \ref{fig3}d,e,f shows the function $\tau(q)$ for the ERAInterim re-analysis for both the NH and the SH, as well as the function $\tau(q)$ for the model runs. All results show a change in the slope of $\tau(q)$ as $q$ passes from negative to positive values. Note that the curve for  geopotential height in the SH exhibits a smaller change than the other curves. The analysis of $\tau (q)$ for the NH zonal mean zonal wind from the ERAInterim re-analysis after a random shuffling of the original time series shows a much weaker change of slope, probably induced by the finite length of the time series (Figure \ref{fig3}d, thick black line).
Overall, the behaviour of the function $\tau(q)$ further confirms the multifractality of  stratospheric variability. To quantify the degree of multifractality we now analyse the singularity spectra of the variability.


\subsection{Singularity Spectra}

\begin{figure}[t]
  \noindent\includegraphics[width=30pc,angle=0]{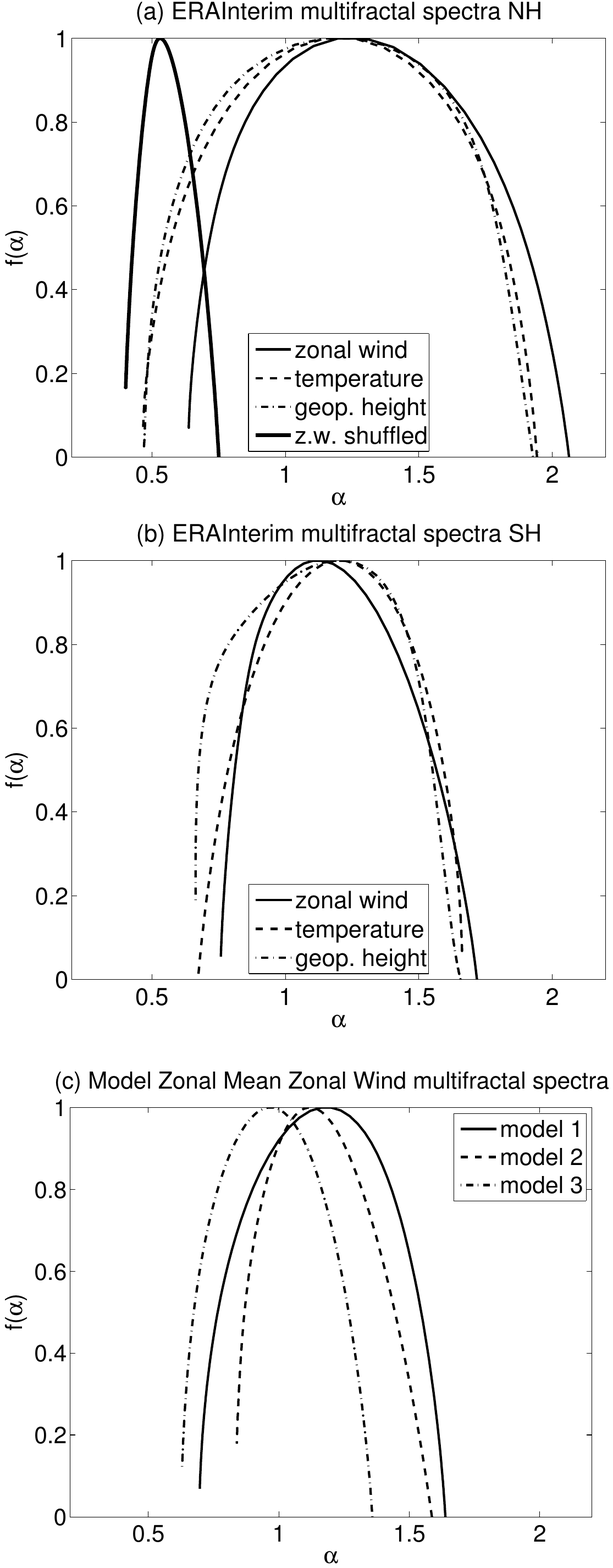}\\
  \caption{Singularity spectra from ERAInterim re-analysis for (a) the Northern Hemisphere and (b) the Southern Hemisphere, for zonal mean zonal wind (solid lines), zonal mean temperature (dashed lines) and zonal mean geopotential height (dot-dashed lines). The think black line in (a) shows the analysis of the Northern Hemisphere zonal mean zonal wind after a random shuffling of the time series. (c) Singularity spectra for the zonal mean zonal wind from Model I (solid line), Model II (dashed line) and Model III (dot-dashed line).}
  \label{fig_multifrac_spectra}
\end{figure}

The singularity spectra $f(\alpha)$ for the ERAInterim re-analysis and for the model runs are shown in Figure \ref{fig_multifrac_spectra}. The width $\Delta \alpha$ of the singularity spectra for the re-analysis reflects the observation that the dynamic range is larger for the NH than the SH, as expected. For the model runs, however, the width of the singularity spectra  is  comparable to the width of $f(\alpha)$ in the SH for Model I, and it is narrower for Model II and Model III. 

The widths of $f(\alpha)$ for the different variables and data sets considered are summarised in Figure \ref{mf_total}a. The results show a coherent picture of larger width in the NH (black dots) than in the SH (white dots), showing thus clearly the correlation between the multifractality of the system and the stronger variability in the NH. In detail, for the NH, the mean width of $f(\alpha)$ is $\Delta \alpha = 1.34 \pm 0.08$. In the SH, instead, the mean width of $f(\alpha)$ is $\Delta \alpha = 0.91 \pm 0.08$. The SH shows also slightly lower values for the width of the singularity spectra for geopotential height. It is in addition curious to observe that the width of the singularity spectra for all variables from the ERA40 re-analysis is slightly smaller than for the other re-analyses. Notice that for all the variables and for all the data sets $\Delta h_{20} / \Delta \alpha \sim 1$, ruling out the presence of spurious multifractality from the requirement $h ( - \infty) - h ( \infty) \sim \alpha |_{q = - \infty} -  \alpha |_{q = \infty}$ \citep{schumann_kantelhardt11}. 

It is  interesting to look at the results for the width of the singularity spectra for the model runs. Model I and Model II show widths that are comparable or slightly smaller than the width for the SH in the re-analysis. In particular, for Model I, the mean width is $\Delta \alpha = 0.93 \pm 0.09$ while for Model II, the mean width is $\Delta \alpha = 0.8 \pm 0.1$.

As pointed out previously, this indicates that the presence of topography (comparing Model II and III) or a seasonal cycle (comparing Model I and III) is not sufficient to trigger a variability with statistical properties comparable to the NH. Model III, represented only by the internal variability of the system, has a much smaller value of the width of the singularity spectra, with mean width $\Delta \alpha = 0.5 \pm 0.2$. To first order, the difference between Model II and Model III is a proxy for the difference between the NH and the SH variability. Qualitatively, this difference is well represented by the fact that the difference in the width of the singularity spectra between the NH and the SH is of the same order as the difference in the width of the singularity spectra between Model II and Model III.

The singularity spectra for the NH zonal mean zonal wind from the ERAInterim re-analysis after a random shuffling of the original time series (Figure \ref{fig_multifrac_spectra}a, thick black line) shows a noisy spectrum centred around a lower value of $\alpha=0.5$ in agreement with the definition for $\alpha$ (\ref{eq:16}) and with the observation that for the randomly shuffled time series $h(q)$ is a constant $h=1/2$. The maximum of the singularity spectra is moved toward smaller values of $\alpha$ also for Model III, in agreement with the observation that the internal variability of the model is close to white noise. The singularity spectra for the randomly shuffled time series shows also a finite, small width $\Delta \alpha = 0.3$ due to the finite length of the time series. The small width of the shuffled time series, for which a zero width would be expected, is thus a measure of the error in the spectral width due to the finite length of the time series, which in this case appears to be smaller than the separation of the width of the spectra between the NH and the SH for the same quantity and re-analysis, that is of $\Delta \alpha (NH) -  \Delta \alpha (SH) \sim 0.5$. Notice that a smaller width of the singularity spectra is observed also for Model III.


\subsection{Sensitivity Analysis with Respect to Location}

While the time series so far have been represented by the location at 10hPa and 60N or 60S, which is often taken to be representative of the extratropical stratosphere, the sensitivity of these results to the chosen latitude and pressure level is tested here. The calculations were repeated for the Northern Hemisphere zonal mean zonal wind from ERAInterim re-analysis at 45N and 10hPa, i.e.~outside of the polar vortex, and at 75N and 10hPa, i.e.~within the polar vortex. The calculations were also repeated at 45N and 75N, both at 10hPa, but at the fixed longitude $\phi=0$. The calculations were also repeated at 10hPa for area weighted averages between 40N-60N, i.e. outside of the polar vortex, and at 10hPa for the area poleward of 60N, i.e.~for the entire polar cap. In order to account for the decreasing area per latitude band towards the pole, the averages have been weighted by $(\cos \theta)^{1/2}$, where $\theta$ is the latitude. Finally, to test for the sensitivity of the results to height, the calculations were repeated in the lower stratosphere at 60N and 50hPa and at 60N and 100hPa.

\begin{figure}[t]
  \noindent\includegraphics[width=30pc,angle=0]{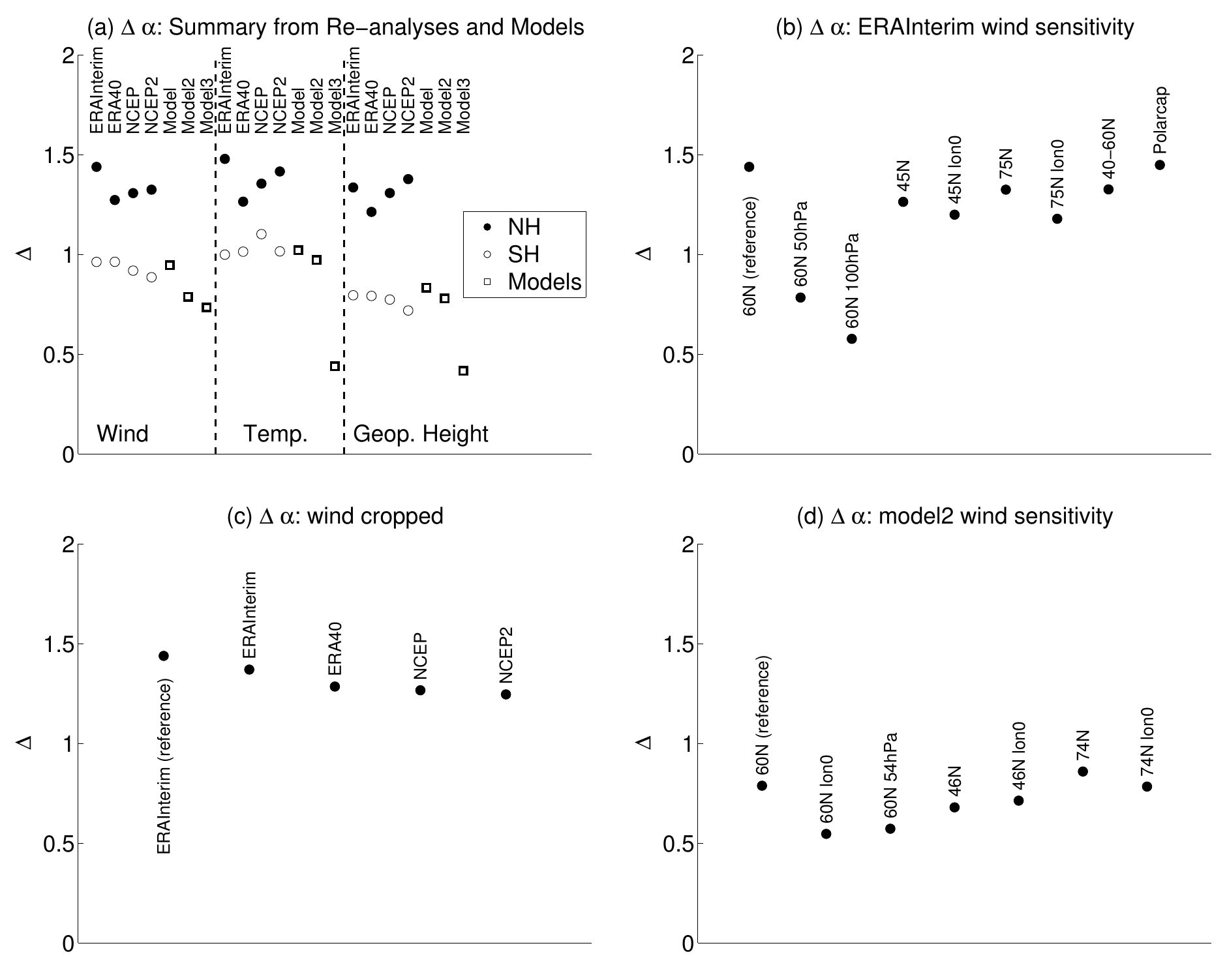}\\
  \caption{Summary of the values for the width $\Delta \alpha$ of the singularity spectra $f(\alpha)$ for (a) all  variables and for all the re-analyses and model runs. Black dots indicate values for the Northern Hemisphere, white dots indicate values for the Southern Hemisphere, and white rectangles indicate values for the model runs; (b) for the Northern Hemisphere zonal mean zonal wind from ERAInterim re-analysis at 60N, 10hPa (reference data set); 60N, 50hPa; 60N, 100hPa; 45N, 10hPa; 45N, 10hPa at longitude 0; 75N, 10hPa; 75N, 10hPa at longitude 0; averaged outside the polar vortex, i.e. between 40N-60N, at 10hPa; and averaged for the polar cap, i.e. north of 60N, at 10hPa; (c) for the Northern Hemisphere zonal mean zonal wind from all re-analysis data sets at 60N, 10hPa, cropped to the common period of 1 Jan 1980 to 31 Dec 2001, in comparison to the reference data set for ERAinterim for 1 Jan 1980 to 31 Dec 2011; (d) for the Model II integration at 60N, 10hPa (reference integration); 60N, 10hPa at longitude 0; 60N, 54hPa; 46N, 10hPa; 46N, 10hPa and longitude 0; 74N, 10hPa;  74N, 10hPa and longitude 0.}\label{mf_total}
\end{figure}

The generalised Hurst exponent for the different sensitivity tests for the Northern Hemisphere zonal mean zonal wind from ERAInterim re-analysis at 60N and 10hPa (not shown) indicate that for all cases $h(-20)-h(0)$ is smaller and $h(0)-h(20)$ is larger than for the reference time series, suggesting that at 60N and 10hPa, small scale fluctuations are more important than for the other time series. In particular, the time series at  60N and 50hPa and at 60N and 100hPa show a much smaller dynamical range. For example,  at  60N and 50hPa results show $\Delta h_{20}= 0.69$. Note that the value is however larger than $\Delta h_{20}$ calculated for the randomly shuffled time series. The smaller variability at 60N and 50hPa and at 60N and 100hPa is also visible in the function $\tau(q)$ which shows a much smaller change of slope as a function of $q$ than the other time series. 
The smaller degree of nonlinearity at  60N and 50hPa and at 60N and 100hPa is strikingly visible in 
Figure \ref{mf_total}c, which shows a summary of the widths of the singularity spectra for the different sensitivity tests. Leaving out the case at 60N and 50hPa and at 60N and 100hPa, the singularity spectra display a mean width of $\Delta \alpha = 1.3 \pm 0.1$, which is comparable to the mean width calculated for the Northern Hemisphere zonal mean zonal wind at 60N and 10hPa from the different data sets previously considered. The width of the singularity spectra at 60N and 50hPa is instead of 0.8 and at 60N and 100hPa is of 0.6, which are comparable, or even less than the value found for the SH. Notice that the value is however larger than the width of the singularity spectra calculated from the randomly shuffled time series. 

These results indicate that the mean width of the singularity spectra at 10hPa is not sensitive to the choice of  latitude, i.e.~on the respective dynamical regions of the NH at 10hPa. This is in agreement with the observation that correlations are fairly strong across the entire hemisphere as they
are driven by the zonally symmetric responses to the wave driving, which on
shorter timescales are characterised by large meridional length scales (e.g. \citet{plumb1982zonally}). The agreement is true also for the time series taken outside and within the polar vortex but at fixed longitude $\phi=0$. In this case in fact, the width of the singularity spectra appears to be slightly narrower than the results from the analysis of the zonal mean, however it does not show significative differences between  45N and 75N, indicating that the width of the singularity spectra is not influenced by the different dynamics there, dominated for example by the potential vorticity filamentation. 
The smaller dynamical range at 60N and 50hPa and at at 60N and 100hPa can instead be explained by the smaller variability at 50hPa and at 100hPa, associated with stronger wave breaking at higher altitudes and by the fact that the radiative timescales become longer lower in the stratosphere. Note that this process is not well represented in the model, which indeed shows a smaller width of the singularity spectra for all the integrations. This is visible in Figure \ref{mf_total}d, which shows the sensitivity of the Model II integration at 60N, 10hPa; 60N, 10hPa at longitude 0; 60N, 54hPa; 46N, 10hPa; 46N, 10hPa and longitude 0; 74N, 10hPa;  74N, 10hPa and longitude 0. Results show that the width of the singularity spectra is consistent for all time series, with a mean width of $\Delta \alpha = 0.7 \pm 0.1$, which confirms the lack of sensitivity of the model integration to short time variability even at different heights. Also in the Model II integration, the width of the singularity spectra does not depend on the choice of taking a fixed longitude.
These results, together with the result of the different width of the singularity spectra between the NH and the SH, show the ability of this method to quantify the variability of the system.

\subsection{Sensitivity Analysis with Respect to the Time Frame}

Another sensitivity test has been performed with respect to the length of the re-analysis time series. As all re-analysis time series have different start and end dates and therefore different lengths, in this experiment the reference time series of zonal mean zonal wind at 60N and 10hPa has been limited to the time period that is common to all re-analysis data sets, i.e.~1 Jan 1980 to 31 Dec 2001. In addition, limiting the data sets to this common time period limits the bias in the analysis that arises from using different methods to assimilating observations into the re-analysis before and after the start of the satellite era in 1979. Limiting the time series to a common period after the start of the satellite era therefore makes the re-analysis data sets more comparable, 
while differences will remain between the different re-analyses as they are produced using different observations, models, and data assimilation techniques, which will constitute the main difference in the findings between the different re-analyses. 

The results for the Hurst exponent, the function $\tau(q)$ (not shown) and the singularity spectra (Figure \ref{mf_total}c) show only a small sensitivity of the results to limiting the time series to the satellite era. In particular the analysis of the Hurst exponent shows that the re-analyses differ, even if only slightly, mainly for negative values of $q$, i.e. in their ability to capture small fluctuations, and agree instead in the large scale variability of the system.

The results for the width $\Delta \alpha$ are reported in Figure \ref{mf_total}c. The full ERAinterim time series at 10hPa and 60N is used as a reference and then compared to the width obtained from the different re-analysis data sets limited to the common time period. The width is slightly smaller for all re-analysis data sets as compared to the reference data set. The different data sets have a mean value of $\Delta \alpha = 1.32 \pm 0.08$, which is in agreement with the mean value calculated for the original time series for the NH. The difference between the data sets lies within the error range of $\Delta \alpha = 0.3$ calculated from the random shuffling of the original time series, and is much smaller than the difference in width between the NH and SH or the re-analysis and the idealised model.


\section{Summary and Discussion}

\label{Discussion}
The stratospheric variability was studied for both the NH and SH searching for signs of multifractal variability in the system, using both data derived from re-analysis data sets and from idealised dynamical core model runs. The NH variability is shown to possess a multifractal nature for time scales shorter than about one year, as seen from the analysis of the fluctuation functions. For these time scales, the variability scales in time with a power law close to $3/2$, corresponding to a red spectrum, as observed by BD14a. For longer time scales, the NH variability becomes monofractal and scales in time with a power law closer to $1/2$, corresponding to white noise. The immediate consequence of these results is that the NH stratospheric variability loses its predictability for time scales longer than about one year, time scales below which variability can be predicted largely from the different variability observed during the seasonal cycle.  

Other impacts that influence predictability in the re-analysis data such as the Quasi-Biennial Oscillation, the El Ni\~no - Southern Oscillation, or volcanic eruptions, are not represented in the idealised model runs. Several of these forcings would be expected to impact the stratosphere via teleconnections on annual or longer time scales, i.e.~influencing the year-to-year variability in the re-analysis data. While the limitations of the model integrations to correctly represent the high frequency variability was already observed studying the frequency spectra in BD 14a,b, at longer time scales, the model tends to more realistically represent the stratospheric variability. This indicates that while remote connections from external predictors will yield improved predictability for a particular winter, as has been shown for the QBO in e.g.~\citet{Scaifeetal2014} and for El Ni\~no in e.g.~\citet{InesonScaife2008}, \citet{ButlerPolvani2011}, \citet{Domeisenetal2015}, the overall range of the variability is already present in the model runs in the absence of these influences, and is characterised by a single (i.e.~monofractal) scaling exponent.

The SH variability shows a similar transition from multifractal to monofractal variability at annual time scales, but with a power law closer to $1$, corresponding to the observed frequency spectrum with a $-1$ slope for time scales shorter than one year. As discussed in BD14b, systems with frequency spectra exhibiting a $-1$ slope are characterised by memory effects, which are associated with the smaller variability of the polar vortex in the SH. The SH also exhibits a narrower dynamical range in multifractality than the NH, due to its weaker variability. These results are found not to be sensitive to the latitude where the time series is captured, and all re-analysis data sets perform similarly when limited to the common time period from 1 Jan 1980 to 31 Dec 2001. 

On the other hand, a non-trivial scaling in (\ref{eq:s1}) could imply the presence of long range correlations, which would instead enhance predictability rather than reducing it. 
The enhancement of the predictability by non-trivial scaling is however dependent on the model considered. 
The general problem of predictability in a multifractal system must include at least two effects \citep{schertzer2004space}: (a)  In deterministic chaotic systems the initial growth of disturbances, which determines the predictability of a system, undergoes a classical exponential growth, with a characteristic time scale given by the Lyapunov exponent. For complex scaling, i.e.~multifractal systems, however, the growth of fluctuations instead follows power laws, for which there are no characteristic time scales and which are characterised by an infinite number of exponents. (b) Multifractality allows for intermittency, and intermittent events are responsible for the loss of information. Further studies will be required to study this effect in the system  considered here, and the way will probably have to go through the formulation of a stochastic model.

The width of the singularity spectra is consistently narrower for the SH than for the NH, in agreement with the results found from the fluctuation analysis. It is interesting to compare the width of the singularity spectra from the re-analyses to the width obtained from the numerical integrations: the latter show a width of the singularity spectra that is comparable to the SH or narrower, indicating that the idealised model is not able to fully capture, at least in statistical terms, the variability of the stratosphere found in re-analysis data. While the longer term variability is comparably well captured in the idealised model runs by the inclusion of large-scale topography and the seasonal cycle, shorter term variability is not adequately represented. Given the very idealised setup of the model runs and the wide use of these runs with the included forcing, a confirmation of the good representation of the large-scale dynamics is re-assuring, while the representation of shorter term variability is found to be limited by the coarser resolution and the corresponding lack of small-scale effects and forcing. Nevertheless, the shortcoming of the models for shorter term variability is intriguing. Future work will be required to further analyse the limitations in the variability of idealised models and more comprehensive model integrations. Comparing the predictability from the methods presented here as well as from other methods, such as ensemble prediction, will elucidate the elements controlling the predictability of a system, and the elements that fall short of giving a variability statistically closer to the re-analyses. 


\section*{Data accessibility}
ERA40 and ERAinterim data have been obtained from the ECMWF Data Server. NCEP re-analysis data was obtained from the ESRL Data Server. The data from the model integrations can be obtained from D. I.V. Domeisen, GEOMAR Helmholtz Centre for Ocean Research Kiel, Germany, E-Mail: ddomeisen@geomar.de. 

\section*{Competing interests}
The authors have no competing interests.

\section*{Authors' contributions}
G. Badin performed the MF-DFA analysis, plotted the figures, and wrote the manuscript draft. D. Domeisen performed the model runs, prepared the input data from the model and reanalysis data, and drafted the sections on stratospheric variability and data. Both authors contributed to the revisions of the manuscript and gave final approval for publication. 

\section*{Funding statement}
G. Badin is funded through U. Hamburg. D. Domeisen is funded through GEOMAR Helmholtz Centre for Ocean Research Kiel / University of Kiel, Germany. 

\section*{Acknowledgements}
The authors would like to thank the anonymous referees for comments that helped to improve the manuscript. 

\ifthenelse{\boolean{dc}}
{}
{\clearpage}
\bibliographystyle{ametsoc}


\clearpage


\clearpage

\clearpage

\clearpage

\end{document}